\documentclass[conference]{IEEEtran}
\IEEEoverridecommandlockouts
\usepackage{cite}
\usepackage{amsmath,amssymb,amsfonts}
\usepackage{algorithmic}
\usepackage{graphicx}
\usepackage{textcomp}
\usepackage{xcolor}
\usepackage{tabularx}
\usepackage{booktabs}
\usepackage{enumitem}
\usepackage{xcolor}
\usepackage{fontawesome5}
\usepackage{tikz}
\usetikzlibrary{
  arrows.meta, positioning, calc, fit, backgrounds,
  shapes.geometric, shapes.misc
}
\usepackage{pgfplots}
\pgfplotsset{compat=1.18}
\usepackage{stfloats}
\begin{document}

\newcommand{\yb}[1]{\textcolor{red}{\small\textbf{[YB]} #1 $\triangleleft$}}

\newcommand{\yash}[1]{\textcolor{blue}{\small\textbf{[YASH]} #1 $\triangleleft$}}

\title{FacProcessTwin: An LLM-Based System\\for Process Twin Development}

\author{
\IEEEauthorblockN{Yash Pulse, Yong-Bin Kang, Abhik Banerjee, Prem Prakash Jayaraman}
\IEEEauthorblockA{
Swinburne University of Technology\\
Melbourne, Australia\\
\{ypulse, ykang, abanerjee, pjayaraman\}@swin.edu.au
}
}

\maketitle

\begin{abstract}
Process twins provide real-time representations of entire production processes. By capturing how process steps interact, rather than monitoring a single machine in isolation as an asset-based digital twin does, they have the potential to drive efficiency gains across the whole process.
However, developing a process twin is costly. It requires accurately modelling the entire production process: its process steps, the equipment and product-specific settings each step uses, and its process variations. The resulting model must then be bound to live operational data. 
We present \textit{FacProcessTwin}, a system that leverages a large language model (LLM) to reduce this development time, building a process twin from a plant's process documentation and natural-language input from an operator.
\textit{FacProcessTwin} generates this complete process model and then automatically binds its process steps to live operational data. 
The generated model and its data bindings are rendered as an interactive process diagram through which manufacturing personnel can monitor and correct the system's autonomous decisions, such as resolving uncertainty at safety-critical binding steps.
We evaluate \textit{FacProcessTwin} through a real-world case study of an Australian food manufacturer, covering 16 production process flows that span chilled, frozen, and aseptic shelf-stable product categories and include process variations within the same product.
The results show that \textit{FacProcessTwin} generates these process models accurately (a mean F1 of 95.2\% against ground truth) and builds each twin in roughly a sixth of the manual time.
Its human-in-the-loop governance then keeps the safety-critical bindings correct: at ambiguous tags where a single-pass baseline silently mis-binds 75.0\% of the time, \textit{FacProcessTwin} defers to the operator and mis-binds none. 
Together, these results show that \textit{FacProcessTwin} makes process twin development practical: it builds accurate twins directly from documentation with far less time and manual effort, while keeping the safety-critical bindings under human oversight.
\end{abstract}


\section{Introduction}
\label{sec:intro}

Digital twins are increasingly deployed on factory floors~\cite{fuller2020digital}. By keeping a virtual model in sync with a physical entity, they enable live monitoring and support applications such as fault prediction, yield forecasting, and production analytics~\cite{tao2019five}. Most such twins, however, are asset-based, modelling a single machine, or sensor in isolation. A \emph{process twin} instead provides a real-time representation of an entire production process (each step, its equipment, and its product-specific settings) and so has the potential to drive efficiency gains across the whole process rather than at a single machine~\cite{verboven2020digital, tancredi2024food}.

While a process twin offers these benefits, developing one is costly and time-consuming, demanding substantial manual effort across two tasks. The first is to model the production process: its process steps, the equipment and product-specific settings each step uses, and its process variations. Much of this information is not written down but held tacitly by operators, which makes the process hard to model: a plant producing food products like sauces, for instance, may document its food-safety steps at a high level while leaving details such as when to change a machine setting to individual operators~\cite{verboven2020digital, tancredi2024food}. The second task is to bind the resulting process model to live operational data, attaching machine settings and parameters to their process steps. Neither task is a one-off: both recur whenever the production process changes, and it is this recurring manual effort that makes process twin development costly and time-consuming~\cite{SU2025841}.

Prior work automates parts of this effort but leaves a gap at each task. One approach generates a virtual process model (typically a simulation) from a formal engineering description that must already exist in machine-readable form, such as a piping and instrumentation diagram (P\&ID)~\cite{martinez2018automatic, sierla2020towards} or an AutomationML model~\cite{alexopoulos2025model}. Most plants do not maintain such inputs, and the output is a model rather than a twin bound to live data. Another applies large language model (LLM) agents to \emph{operate} an existing twin for control or fault handling~\cite{xia2025control, gill2025leveraging} rather than to develop one. A third generates twin information models from equipment datasheets~\cite{xia2024generation} but leaves the resulting data bindings unvalidated. No prior system develops a complete, data-bound twin from the narrative documentation a plant already maintains. Doing so has two demanding parts: recovering an accurate process model from loose, unstructured prose and tables, and binding that model to live data without error, since an agent acting without oversight will guess at an ambiguous tag precisely where a wrong binding is a safety hazard.

To develop a process twin directly from a plant's process documentation, we present \textit{FacProcessTwin}, a system that leverages an LLM to model the production process from that documentation and an operator's natural-language input, and to bind the resulting model to live operational data. To keep development safe, \textit{FacProcessTwin} places a human in the loop at the binding step: rather than guessing at an ambiguous, safety-critical binding, it pauses and asks an operator. The process model and its live bindings are presented as an interactive process diagram that manufacturing personnel use to review and correct the system's decisions.

We evaluate \textit{FacProcessTwin} on a real food-manufacturing line whose documentation specifies 16 production process flows (across chilled, frozen, and aseptic shelf-stable categories, with small within-product variations), each with a known-correct ground-truth model, and we test robustness across four LLM backbones.

Specifically, this paper makes two contributions:
\begin{enumerate}[leftmargin=1.4em,itemsep=2pt,topsep=2pt,parsep=0pt]
    \item We propose \textit{FacProcessTwin}, an end-to-end system that develops a process twin from process documentation and natural-language input, automating both process modelling and the binding of process steps to live machine data, with a human-in-the-loop step that refers ambiguous, safety-critical bindings to an operator.
    \item We evaluate \textit{FacProcessTwin} on 16 real-world production process flows from a mid-sized Australian food manufacturer, measuring the accuracy of the generated process model, its behaviour at safety-critical binding decisions, and its development effort relative to manual modelling.
\end{enumerate}

The remainder of this paper reviews related work (Section~\ref{sec:related}), presents \textit{FacProcessTwin} and its safeguards (Section~\ref{sec:overview}), reports the evaluation and results (Sections~\ref{sec:evaluation}, and concludes (Section~\ref{sec:conclusion}).

\section{Related Work}
\label{sec:related}


Work most relevant to \textit{FacProcessTwin} falls into three areas: automated twin generation, LLM agents in manufacturing, and representations and data binding.

\textbf{Automated Twin Generation:}
Prior work generates simulation-based twins from P\&IDs and 3D~CAD~\cite{martinez2018automatic, sierla2020towards}, and recent work extends this to AutomationML-driven generation with generative-AI components~\cite{alexopoulos2025model}. Semantic, simulator-backed ``cyber twins'' have also been proposed for Industry~4.0~\cite{bamunuarachchi2021cybertwins} and extended into digital-manufacturing lifecycle platforms~\cite{gunaratne2026digital}. These approaches consume structured or graphical engineering artefacts rather than narrative procedures, and produce simulation models or require hand engineering to reach a data-connected twin. In contrast, \textit{FacProcessTwin} parses process documentation (prose and tables) and produces a twin bound to live data from Open Platform Communications Unified Architecture (OPC~UA) servers

\textbf{LLM Agents in Manufacturing:}
LLM agents have been applied to plan and control modular production from twin descriptions~\cite{xia2025control}. \textit{AROMA-GPT}, the closest system to ours, couples an LLM to a physics-based reactor twin over bidirectional OPC~UA with a human-in-the-loop interface~\cite{ndum2026large}. Other work pairs monitoring and action agents with a twin to recover a process module after faults~\cite{gill2025leveraging}, while \textit{AssetOpsBench}~\cite{patel2025assetopsbench} and the \textit{SmartPilot} copilots~\cite{shyalika2025smartpilot} benchmark and coordinate such agents. Across these systems, the twin is assumed to already exist: they operate or diagnose it, but do not develop it or validate its data bindings from documentation.

\textbf{Representations and Data Binding:}
OPC~UA exposes plant data as a node graph: an address space of objects, relationships, and metadata, not mere tag values. The Asset Administration Shell (AAS) standardises twin information models, and has been generated with LLM assistance from equipment datasheets~\cite{xia2024generation}.

Across this prior work, two gaps recur: twin generation relies on structured engineering artefacts rather than the narrative documentation plants maintain, and the LLM--twin literature operates existing twins rather than developing them or safeguarding their bindings. \textit{FacProcessTwin} addresses both: it develops a live, OPC~UA--bound process twin from process documentation under a human-in-the-loop workflow.

\section{FacProcessTwin}
\label{sec:overview}

\textit{FacProcessTwin} rests on a single design principle: the information needed to build a process twin already exists in a plant, just not in a usable form. Process knowledge is spread across documented procedures and the tacit know-how of operators; the production process of an industrial mixing plant, for instance, was captured partly in recipe documentation and partly in operators' tacit knowledge~\cite{LoDiSA24Paper}. Rather than requiring a pre-built, machine-readable model, \textit{FacProcessTwin} uses an LLM to recover this dispersed, largely unstructured knowledge through natural language: it reads the plant's documents alongside an operator's answers, models the production process, and binds that model to live machine data.

\textit{FacProcessTwin} differs from earlier work in three ways. First, it \emph{develops} a twin rather than operating one that already exists. Second, it builds that twin from narrative process documentation (prose and tables) and an operator's natural-language input, rather than from structured engineering artefacts such as P\&IDs or AutomationML models. Third, and central to its design, it runs under \emph{human-in-the-loop governance}. The LLM is advisory only: it reads and proposes, deterministic tools carry out every change, and the operator confirms or resolves any decision that is ambiguous or safety-critical, so the system pauses for a human rather than guessing. This governance acts at three checkpoints: choosing the product flow, approving the drawn layout, and clarifying an ambiguous binding. We detail how it performs at Steps~2 and~4.

\begin{figure*}[t]
\centering
\includegraphics[width=0.91\textwidth]{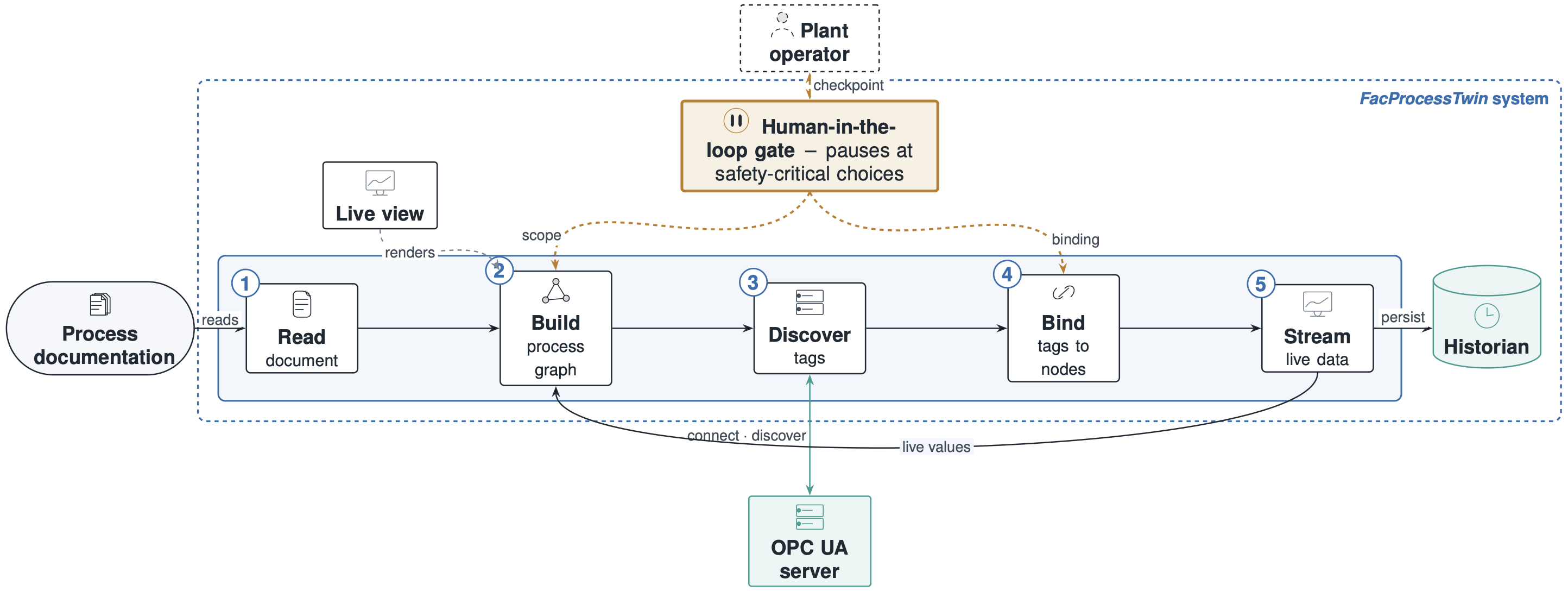}
\caption{Architecture of \textit{FacProcessTwin}. The numbered stages (1--5) are executed inside the human-in-the-loop LLM agent and its tool set (shaded region; Table~\ref{tab:tools}). The dashed boundary encloses the system; the process documentation, operator, and OPC~UA server lie outside it.}
\label{architecture}
\end{figure*}

The system runs as a single pipeline (Fig.~\ref{architecture}): (1) it reads the chosen document; (2) it extracts the process steps and draws them on a canvas; (3) it connects to the plant's OPC~UA servers and lists the tags they expose; (4) it binds each tag to the matching step; and (5) it streams live values onto the canvas and stores them. Each stage runs through a small set of self-contained, deterministic tools (Table~\ref{tab:tools}) that the agent invokes but never bypasses, and the design is extensible: a new document format or server protocol is added as another tool, without changing the loop or its safeguards. We present how each step performs below, using one aseptic-pur\'ee flow as a running example.

\begin{table}[t]
\centering
\caption{Tools the agent can call, by stage. The model selects a call; the tool executes it.}
\label{tab:tools}
\small
\begin{tabularx}{\columnwidth}{@{}lX@{}}
\toprule
\textbf{Stage} & \textbf{Tools} \\
\midrule
Ingestion & read document (PDF/DOCX); extract text and tables; recover candidate product flows \\
Process graph & generate nodes and edges; compute layout; persist to the EAV store \\
Connection \& discovery & connect to an OPC~UA server; list tags, with vendor-filter, direct-read, and identifier-repair fallbacks \\
Binding & create a binding and subscribe; verify against recorded state; write to the historian \\
Interaction & ask a question; request a tag clarification \\
\bottomrule
\end{tabularx}
\end{table}

\textbf{1. Read Document:}
Development begins from the plant's process documentation, the primary source of the process model. This documentation is free-form (prose and tables in PDF or DOCX, written for people, not machines) and usually describes more than one product. A reading tool extracts its text and tables and passes them to the model, which recovers the candidate product flows. Because one document typically covers many products (16 in our evaluation), the system surfaces the candidates it finds and, at the first human-in-the-loop checkpoint, asks the operator which flow to develop.

\textbf{2. Build Process Graph:}
This step builds the twin's \emph{process model}, represented as a \emph{process graph} of \emph{nodes} (tanks, pumps, sensors, or generic process steps) joined by directed \emph{edges} that show how product flows between steps. Once the operator has chosen a flow (Step~1), that flow becomes an ordered list of steps, each carrying its \emph{control-point} label, a designation from the plant's Hazard Analysis and Critical Control Point (HACCP) plan that marks safety-relevant steps. In the running example, the steps run from raw-material inspection and washing, through cutting and cooking, to form-fill-seal packaging, ultra-high-temperature (UHT) sterilisation, and aseptic filling.

The system then draws these nodes and edges on a canvas. Because letting the model place them risks overlapping boxes or wrong links, the layout is computed by fixed code rather than by the model. Human-in-the-loop governance now takes effect: the drawn graph is shown to the operator, and development proceeds to binding only once the operator approves it. The graph is stored in a flexible entity--attribute--value (EAV) format rather than fixed tables, so its shape can vary from plant to plant without rebuilding the database, which is necessary because process documents rarely use the same equipment names.

\textbf{3. Discover Tags:}
Before binding, \textit{FacProcessTwin} must find out which data each machine exposes. It connects to the plant's OPC~UA servers, the common standard for live machine data, and lists the \emph{tags} they publish, each tag being a single live value such as a temperature reading. The difficulty is practical, not conceptual: industrial servers list their data in vendor-specific ways, so a normal request can return empty, as it does for the Siemens~S7 and Prosys servers in our setup. A discovery tool handles this with three fallbacks: it retries with the right vendor filter; it reads tags directly when a listing comes back empty; and, after a server restart, it repairs changed identifiers and reconnects once the link returns.

\textbf{4. Bind Tags to Nodes:}
\label{sec:method:tagtonode}
Binding links the process model to live data: a \emph{binding} ties an OPC~UA tag to a node. This is where acting without guidance is most dangerous: a confident but wrong binding, say a fill-temperature reading attached to the wrong step, quietly corrupts the twin and may not surface until weeks into production. Human-in-the-loop governance is therefore strongest here: when a tag clearly belongs to one node, \textit{FacProcessTwin} creates the binding and starts listening for updates; when the target is unclear, it stops and asks the operator rather than guessing. In the running example, the UHT server exposes two temperature tags (here labelled T1 and T2), and either could fit the sterilisation step; because that step is a critical control point, the system refuses to choose and asks the operator which tag is the holding-tube temperature. Any change that would overwrite existing data needs explicit confirmation, and a final check compares the twin against recorded values before it goes live.

\textbf{5. Stream Live Data:}
Once bindings are confirmed, the twin goes live: \textit{FacProcessTwin} subscribes to each bound tag, streams its current value onto the matching node so the model tracks the running process in real time, and writes each reading to the plant's historian. These readings flowing back onto the diagram close the loop between the physical process and its twin, giving operators a single view of the running process and the bindings that feed it.

Two properties hold across the whole pipeline (Fig.~\ref{architecture}). First, the LLM's free judgement enters at only two points, recovering the list of steps and proposing bindings, while the runtime drives it in a checked loop with a fixed step limit of 40 (Section~\ref{sec:evaluation:method}) and deterministic tools perform every change; a LLM error therefore surfaces as a missing step or a deferred binding, caught at a checkpoint, never as a silently corrupted graph. Second, \textit{FacProcessTwin} depends on no particular model or vendor: it works with any LLM and any OPC~UA server, a claim we test directly in Section~\ref{sec:results}. In Fig.~\ref{architecture}, these five steps and their tools run inside the human-in-the-loop agent (the shaded region), while the process documentation, the operator, and the OPC~UA servers lie outside the system boundary.
\section{Evaluation}
\label{sec:evaluation}

\subsection{Evaluation Method}
\label{sec:evaluation:method}

Our evaluation asks two things: how accurately and efficiently \textit{FacProcessTwin} develops a process twin from documentation, and whether its human-in-the-loop governance keeps the safety-critical bindings correct. We assess construction along three axes: the fidelity of the recovered process model, the accuracy of the tag mappings, and the development effort against manual modelling. We then test the governance directly with a human-in-the-loop ablation, measuring whether it prevents the hazardous mis-bindings a single-pass baseline makes when a tag at a critical control point is ambiguous, without degrading accuracy on the bindings an agent could safely make alone. All are reported under one set of metric names, reused in the results that follow.

\textbf{Dataset:}
Our dataset comprises the process documentation of a mid-sized Australian food manufacturer, in the form of a HACCP standard operating procedure (SOP). The SOP specifies a library of 26 coded process steps and 16 production process flows, each given as an explicit, ordered step sequence with control-point annotations that serve as ground truth: a known-correct process model against which \textit{FacProcessTwin}'s output is scored. The 16 flows span chilled (7), frozen (5), and aseptic shelf-stable (4) categories; 13 are smooth and 3 particulate, and together they instantiate the 26 library steps 225 times (11--18 per flow, mean 14). The dataset is deliberately more than a single case: prior LLM-twin studies typically report one or a few demonstrations~\cite{xia2024generation,ndum2026large,gill2025leveraging}, whereas this SOP supplies 16 flows over one consistent step vocabulary, allowing both per-flow scoring and an aggregate. Because these flows share one document and vocabulary, the aggregate measures robustness within a single SOP rather than generality across sites (Section~\ref{sec:conclusion}). 

\textbf{Metrics:}
Process structure and tag mapping are assessed with several measures, computed per flow and then aggregated. \emph{Topology F1} is the harmonic mean of precision and recall over the extracted steps. \emph{Sequence accuracy} is one minus the normalised edit distance between the extracted and ground-truth step sequences $S_{\mathrm{ext}}$ and $S_{\mathrm{gt}}$,
\begin{equation}
\mathrm{SeqAcc} = 1 - \frac{\operatorname{lev}\!\left(S_{\mathrm{ext}}, S_{\mathrm{gt}}\right)}{\max\!\left(|S_{\mathrm{ext}}|,\ |S_{\mathrm{gt}}|\right)},
\label{eq:seqacc}
\end{equation}
where $\operatorname{lev}(\cdot,\cdot)$ is the Levenshtein edit distance and $|\cdot|$ is sequence length; the score lies in $[0,1]$, equalling $1$ on an exact match, with the exact-match rate reported alongside. \emph{Control-point accuracy} is the proportion of control-point labels (CCP, OCP, QCP) correctly carried, a safety-relevant signal. \emph{Mapping precision and recall} are measured against a hand-labelled tag-to-node ground truth. We then isolate the effect of the human-in-the-loop governance with two measures that together avoid scoring the policy against itself. The \emph{false-mapping rate} is the fraction of genuinely ambiguous tags an agent binds silently and incorrectly rather than raising for clarification; this is near zero by design for the human-in-the-loop agent, so the informative quantity is the single-pass baseline's error rate on the same tags. \emph{Unambiguous-tag accuracy} is each agent's accuracy on the non-ambiguous tags, reported so that any reduction in false mappings can be checked against overall accuracy.

\textbf{Evaluation Settings:}
All experiments fix the agent backbone to a single model (Gemini~3.1 Flash-Lite), so that score differences reflect the method rather than the model; robustness across three further models (GPT-5-mini, DeepSeek-V4-Flash, and MiniMax-M2.7) is examined separately in the results below. The agent runs with a step budget of 40. Live machine data is served over OPC~UA by two servers: a UHT/pasteuriser server exposing 15 tags and a cooker server exposing 4 tags across two cooker units (each flow uses one). The 4 aseptic flows bind both servers; the other 12, having no pasteurisation step, bind the cooker only. The manual baseline configures the same twin and tags by hand under identical timed conditions; we treat this single-builder figure as indicative of order-of-magnitude effort, not a population estimate (Section~\ref{sec:conclusion}). For each flow, \textit{FacProcessTwin} ingests the SOP, the operator confirms scope, and the agent develops and binds the twin, recording \textit{time-to-twin} and \textit{intervention count}; the artefact is then compared against ground truth. For the human-in-the-loop ablation, we re-run the same flows with a single-pass baseline that omits the staged checkpoints and disables the clarification policy, recording how often it binds genuinely ambiguous tags silently and incorrectly.
\subsection{Evaluation Results}
\label{sec:results}

Across the 16 flows, \textit{FacProcessTwin} develops a faithful, data-bound twin at a fraction of the manual time, and its human-in-the-loop governance keeps the safety-critical bindings correct where the single-pass baseline fails. We report the evidence in turn: that the twin is accurate, that its bindings are safe, and what it costs to build, under the metric names introduced above.

\begin{figure}[t]
\vspace*{6pt}
\centering
\resizebox{0.88\columnwidth}{!}{%
  \definecolor{Signal}{HTML}{2A6FB0}
\begin{tikzpicture}
\begin{axis}[
  xbar, bar width=11pt,
  width=\columnwidth, height=4.4cm,
  xmin=0, xmax=116,
  xlabel={Score (\%)},
  symbolic y coords={Mapping precision, Mapping recall, Control-point acc., Topology recall, Topology precision, Topology F1},
  ytick=data,
  tick label style={font=\small}, label style={font=\small},
  nodes near coords, nodes near coords align={horizontal},
  every node near coord/.append style={font=\footnotesize},
  axis y line=left, axis x line=bottom,
  enlarge y limits=0.12,
]
\addplot[fill=Signal, draw=Signal!60!black] coordinates {
  (95.2,{Topology F1}) (95.4,{Topology precision}) (95.0,{Topology recall})
  (96.4,{Control-point acc.}) (100,{Mapping recall}) (74.2,{Mapping precision})
};
\end{axis}
\end{tikzpicture}
}
\caption{Topology and mapping fidelity (per-flow means).}
\label{fig:fidelity}
\end{figure}
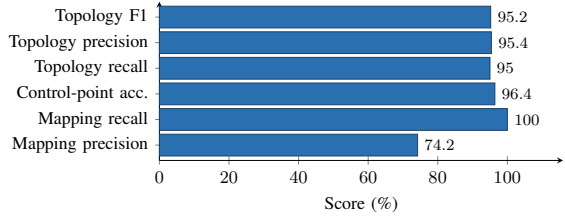

\textbf{Topology Fidelity:}
\textit{FacProcessTwin} recovers the process structure accurately (Fig.~\ref{fig:fidelity}). Of 225 ground-truth steps it recovered 214 and introduced 10 invalid nodes, a pooled \emph{topology F1} of 95.3\% (per-flow mean 95.2\%, range 85.7--100\%), with \emph{topology precision} 95.4\% and \emph{topology recall} 95.0\% closely matched; the agent neither over- nor under-generates steps systematically. Because the task is recovery rather than discovery (the SOP gives the sequence), the residual error is small and concentrated rather than scattered: two preparation steps (peeling, washing) were dropped on one frozen flow, and one aseptic filler was over-segmented into two nodes. No flow fell below 85.7\% F1.

\emph{Sequence accuracy} averages 90.2\%, with exact reconstruction of the step order on 8 of 16 flows. Most flows are at or near exact; a single low outlier (15.4\%) pulls the mean down: a flow that reached the 40-step budget before completing its ordering, a runtime limit rather than a reasoning failure. \emph{Control-point accuracy}, the safety-relevant signal that the control-point designations (CCP, OCP, QCP) survive extraction, is 96.4\%: on three flows every step was recovered but one or two control-point classes were mislabelled, exactly the kind of slip a deployment would route to operator review.

\textbf{Mapping Fidelity:}
Binding is complete but deliberately conservative (Fig.~\ref{fig:fidelity}). \emph{Mapping recall} is 100\%: all 92 required tags are placed on the correct node, so nothing the twin needs is missed. \emph{Mapping precision} is 74.2\%, the one metric below 90\%, from a single identifiable cause: the agent created 32 extra bindings, all to a second cooker unit the server exposes but these flows do not use. The SOP never states which of the two cooker units a flow runs on, so the shortfall is missing information in the document, not faulty reasoning by the agent. This unit-level ambiguity escapes the binding check, which asks which tag fits a step, not which physical unit a flow uses: each cooker-temperature tag is a valid match for the cook step, so the agent binds both. The governance instead has its clean test on a separate, narrower set — the 20 tags where two candidates compete for the same step, which the agent declined to bind and deferred to the operator.

\textbf{Human-in-the-Loop Governance:}
This experiment isolates the effect of the governance, answering the second of our two evaluation questions. On the 20 ambiguous tags, the \emph{false-mapping rate} is decisive (Table~\ref{tab:hitl_ablation}): The single-pass baseline binds all 20 without asking and gets 15 wrong (a 75.0\% error rate), including the holding-tube temperature on the aseptic flows, where it chooses blind between the two temperature tags T1 and T2 (Section~\ref{sec:method:tagtonode}). That step is a critical control point, so a wrong binding there is precisely the failure the governance exists to prevent: it is not a cosmetic error but the safety hazard named in the introduction. With the human-in-the-loop (HITL) governance in place, the same 20 decisions are deferred and the \emph{false-mapping rate} falls to 0\%.

This caution costs nothing on the bindings an agent could make safely: \emph{unambiguous-tag accuracy} is statistically indistinguishable between the variants (90.0\% with the governance against 89.4\% without). The governance does not trade accuracy for safety: it removes the hazardous errors while leaving the easy bindings untouched. The 20 deferrals are real operator effort, whose acceptability we leave to a future operator-acceptance study (Section~\ref{sec:conclusion}); the case for asking is the 75.0\% blind-binding rate it avoids.

\begin{table}[t]
\centering
\caption{Human-in-the-loop ablation across the 16 flows.}
\label{tab:hitl_ablation}
\footnotesize
\setlength{\tabcolsep}{4pt}
\begin{tabular}{lcc}
\toprule
\textbf{Metric} & \textbf{With HITL} & \textbf{Without HITL} \\
\midrule
Ambiguous bound & 0 (policy) & 20 \\
False-mapping rate & 0.0\% & 75.0\% (15/20) \\
Unambiguous-tag accuracy & 90.0\% & 89.4\% \\
\bottomrule
\end{tabular}
\end{table}

\begin{figure}[t]
\centering
\resizebox{0.88\columnwidth}{!}{%
  \definecolor{Signal}{HTML}{2A6FB0}
\begin{tikzpicture}
\begin{axis}[
  xbar, bar width=15pt,
  width=\columnwidth, height=3.0cm,
  ytick={1,2}, yticklabels={\textit{FacProcessTwin}, Manual},
  ymin=0.5, ymax=2.8, xmin=0, xmax=48,
  xlabel={Time per flow (min)},
  tick label style={font=\small}, label style={font=\small},
  nodes near coords, every node near coord/.append style={anchor=south, yshift=7pt, font=\footnotesize},
  axis y line=left, axis x line=bottom,
]
\addplot[fill=Signal, draw=Signal!60!black,
  error bars/.cd, x dir=both, x explicit]
  table[x=v, y=p, x error plus=ep, x error minus=em, row sep=\\]{
  v p ep em \\
  5.2 1 2.3 1.8 \\
  31.8 2 8.3 6.6 \\
  };
\end{axis}
\end{tikzpicture}
}
\caption{Development effort per flow (mean; whiskers show the per-flow range).}
\label{fig:effort}
\end{figure}
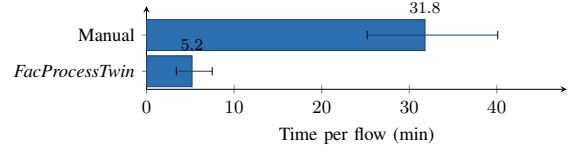

\textbf{Development Effort and Robustness:}
Safety is only useful if development stays practical. Measured by \emph{time-to-twin}, \textit{FacProcessTwin} builds and binds a twin in 5.2 minutes per flow on average (range 3.4--7.5) against 31.8 minutes for the manual baseline (range 25.2--40.1), an 83.6\% reduction and roughly a sixfold speed-up (Fig.~\ref{fig:effort}), at an \emph{intervention count} of two operator questions per flow. The human burden is therefore small, and even the slowest automated run beats the quickest manual one.

The result also holds across LLM backbones: re-running the full evaluation on GPT-5-mini, DeepSeek-V4-Flash, and MiniMax-M2.7 alongside the default Gemini~3.1 Flash-Lite, \emph{topology F1}, \emph{sequence accuracy}, and \emph{time-to-twin} vary by no more than about $\pm$1 percentage point, supporting the model-agnostic claim of Section~\ref{sec:overview}.

\textbf{Findings and Implications:}
Three implications follow. First, developing a process twin from ordinary documentation is feasible at scale rather than merely demonstrable, and its ceiling is set by what the documentation omits: the only sub-90\% metric reflects missing information (which cooker unit a flow uses), pointing to richer data sources such as enterprise resource planning (ERP) and manufacturing execution system (MES) feeds rather than a larger model~\cite{SU2025841}. Second, the human-in-the-loop governance turns the most dangerous failure mode (a silent mis-binding at a critical control point) into a bounded operator cost, which is what makes the approach usable in an operational-technology setting. Third, because quality is set by the policy and prompting rather than the backbone, the system should carry over as models improve; the threat that remains is generality beyond a single SOP and site, which we address in Section~\ref{sec:conclusion}.
\section{Conclusion and Future Work}
\label{sec:conclusion}

This paper set out to close the gap identified in Section~\ref{sec:intro}: no prior system develops a complete process twin from the ordinary documentation a plant already keeps, which demands both recovering an accurate process model from loose prose and binding it to live machine data, the latter safety-critical because an agent acting without oversight will otherwise guess at an ambiguous tag precisely where a wrong binding would be a hazard. \textit{FacProcessTwin} closes this gap: from a plant SOP and an operator's natural-language input, its LLM recovers the process structure and proposes the live-tag bindings, deterministic tools apply them, and, where a binding would be unsafe, the system defers to the operator instead of guessing. The paper makes two contributions: first, a system that develops a complete, data-bound process twin from documentation, with human-in-the-loop governance that converts the LLM's unsafe guesses into operator decisions at safety-critical steps; and second, an evaluation on 16 ground-truth production process flows from a real food-manufacturing HACCP SOP. That evaluation (Section~\ref{sec:results}) shows that \textit{FacProcessTwin} builds a faithful twin from documentation alone (95.2\% F1, 100\% mapping recall) at a sixth of the manual time, while its governance cuts silent, safety-critical mis-bindings from a 75.0\% rate to 0\%.

Two limitations set the agenda. The evaluation rests on a single plant's SOP whose 16 flows share wording and structure, so the results hold within this document rather than across sites; testing \textit{FacProcessTwin} on SOPs from other plants is the first priority. The governance also flags each unsafe decision for the operator rather than resolving it automatically, so the resulting operator effort, and whether it is acceptable in practice, warrants a dedicated operator-acceptance study. Two design directions follow: feeding the agent the production data held in ERP and MES systems would settle bindings the SOP cannot determine, such as which cooker unit a flow uses; and because each capability is a self-contained tool, new document formats or protocols can be added without rebuilding the system. Together these point towards reliable, low-effort process twins for the mid-sized manufacturers this work targets.

\section*{Acknowledgment}

This research was supported in part by the Australian Research Council (ARC) Industrial Transformation Research Hub for Future Digital Manufacturing (DMH), IH230100013

\bibliographystyle{IEEEtran}
\bibliography{references}

\end{document}